\documentclass{article}

\usepackage{amsmath}
\usepackage{hyperref}
\usepackage{graphicx}

\begin{document}

\title{Solving Polynomial Systems with phcpy}

\author{Jasmine Otto\thanks{University of California, 
Santa Cruz, jtotto@ucsc.edu} \and
Angus Forbes\thanks{University of California, Santa Cruz, angus@ucsc.edu} \and
Jan Verschelde\thanks{University of Illinois at Chicago, janv@uic.edu}}

\maketitle

\begin{abstract}
The solutions of a system of polynomials in several variables are often
needed, e.g.: in the design of mechanical systems, and    in phase-space analyses of nonlinear biological dynamics.    Reliable, accurate, and comprehensive numerical solutions are available    through PHCpack, a FOSS package for solving polynomial systems with   homotopy continuation.

This paper explores new developments in phcpy, 
a scripting interface for PHCpack, over the past five years. 
For instance, phcpy is now available online through a JupyterHub server 
featuring Python2, Python3, and SageMath kernels. As small systems are 
solved in real-time by phcpy, they are suitable for interactive exploration
through the notebook interface. 
Meanwhile, phcpy supports GPU parallelization, improving the speed 
and quality of solutions to much larger polynomial systems. 
From various model design and analysis problems in STEM, certain classes 
of polynomial system frequently arise, to which phcpy is well-suited.
\end{abstract}

\section{Introduction}

The Python package phcpy \cite{Ver14} provides an alternative to the
command line executable \texttt{phc} of PHCpack \cite{Ver99} 
to solve polynomial
systems by homotopy continuation methods. In the phcpy interface,
Python scripts replace command line options and text menus,
and data persists in a session without temporary files.
This also makes PHCpack accessible from Jupyter notebooks,
including a JupyterHub server available online \cite{Pascal}.

phcpy takes as input a list of polynomials in several variables, 
with complex-valued floating-point coefficients.
Homotopy methods connect this given system to a 'start system' 
with known solutions.
A homotopy is a family of polynomial systems where one of the variables 
is considered as a parameter.
Polynomial homotopy continuation combines the application of homotopy 
and continuation methods, which extend the convergence of Newton's method 
from local to global, to solve polynomial systems.

Numerical continuation methods track the solution paths, 
depending on the parameter,
originating at the known solutions to the solutions of the given system.
phcpy is also able to represent the numerical irreducible decomposition 
of the system's solution set, which yields 
the \emph{positive dimensional solution sets} containing infinitely 
many points, in addition to the isolated solutions.

The focus of this paper is on the application of new technology
to solve polynomial systems, in particular, cloud computing \cite{BSVY15}
and multicore shared memory parallelism
accelerated with graphics processing units \cite{VY15}.
Our web interface offers phcpy in a SageMath \cite{Sage}, \cite{SJ05} kernel
or in a Python kernel of a Jupyter notebook \cite{Klu16}.

Although phcpy has been released for only five years,
three instances in the research literature of symbolic computation,
geometry and topology, and chemical engineering (respectively)
mention its application to their computations.

\begin{itemize}
\item The number of embeddings of minimally rigid graphs \cite{BELT18}.
\item Roots of Alexander polynomials \cite{CD18}.
\item Critical points of equilibrium problems \cite{SWM16}.
\end{itemize}

The package phcpy is in ongoing development. At the time of writing,
this paper is based on version 0.9.5 of phcpy,
whereas version 0.1.5 was current at the time of \cite{Ver14}.
An example of these changes is that the software described in \cite{SVW03}
was recently parallelized for phcpy \cite{Ver18}.

\subsection{A Scripting Interface for PHCpack}

The mission of phcpy is to bring polynomial homotopy continuation
into Python's computational ecosystem.

The package phcpy wraps the shared object files of a compiled PHCpack,
which makes the methods more accessible without sacrificing their efficiency.
First, the wrapping transfers the implementation of the many available homotopy algorithms in a direct way into Python modules.
Second, we do not sacrifice the efficiency of the compiled code.
Scripts replace the input/output movements and interactions with the user,
but not the computationally intensive algorithms.

Numerical algebraic geometry \cite{SVW05} was introduced in 1995 as a pun on
numerical linear algebra.
PHCpack prototyped the first algorithms to compute
a numerical irreducible decomposition of the solution set
of a polynomial system.
The package phcpy aims to bring the algorithms of numerical algebraic geometry
into the computational ecosystem of Python.

\subsection{Related Software}

PHCpack is one of three FOSS packages for polynomial homotopy computation
currently under development.  Of these, only Bertini 2 \cite{Bertini2.0}
also offers Python bindings, although it is not GPU-accelerated and 
does not export the numerical irreducible decomposition, 
among other differences.
Version 1.4 of Bertini is described in \cite{BHSW13}.

HomotopyContinuation.jl \cite{HCJL} is a standalone package for Julia,
presented at ICMS 2018 \cite{BT18}.

NAG4M2 \cite{NAG4M2} is a package for Macaulay2
(a standard computational algebra system \cite{M2}),
which can also act an interface to PHCpack or Bertini.
As described in \cite{Ley11}, it provided the starting point
for PHCpack's Macaulay2 bindings \cite{GPV13}.

\section{User Interaction}

\subsection{Online Access}

The first area of improvement that phcpy brings is in the interaction 
with the user.

With JupyterHub \cite{JUPH}, we provide online access \cite{Pascal} 
to environments with Python and SageMath kernels pre-installed, 
both featuring phcpy and tutorials on its use (per next section). 
Since Jupyter is language-agnostic, execution environments in several 
dozen languages are possible. Our users can also run code in a Python 
Terminal session. As of the middle of May 2019, our web server has 146 
user accounts, each having access to our JupyterHub instance. 
Our server is available for public use, after creating a free account.

In our first design of a web interface to \texttt{phc}, 
we developed a collection of Python scripts (mediated through HTML forms), 
following common programming patterns \cite{Chu06}. 
This design is described in Chapter 6 of \cite{Yu15}. 
For the user administration of our JupyterHub, we refreshed this first 
web interface, retaining the following architecture.

MySQLdb \cite{MSDB} does the management of user data, including
a) names and encrypted passwords,
b) generic, random folder names to store data files, and
c) file names with polynomial systems they have solved.
With the module smtplib, we defined email exchanges for an automatic
2-step registration process and password recovery protocol.

A custom JupyterHub Authenticator connects to the existing MySQL database
and triggers a SystemdSpawner that isolates the actions of users to separate
processes and logins in generic home folders. 
The email account management prompts were hooked 
to new Tornado RequestHandler instances, which perform account registration
and activation in the database, as well as password recovery and reset. 
Each such route serves HTML forms seamlessly with the JupyterHub interface, 
by extending its Jinja templates.

\subsection{Code Snippets}

Learning a new API is daunting enough without also being a crash course 
in algebraic geometry.  Therefore, the user's manual of phcpy \cite{PHCPY}
begins with a tutorial section using only the blackbox solver 
\texttt{phcpy.solver.solve(system, ...)}. 
In this API, \texttt{system} is a list of strings representing polynomials, 
terminated by semicolons, and containing as many variables as equations.

\begin{figure}[h]
{\includegraphics[width=\columnwidth]{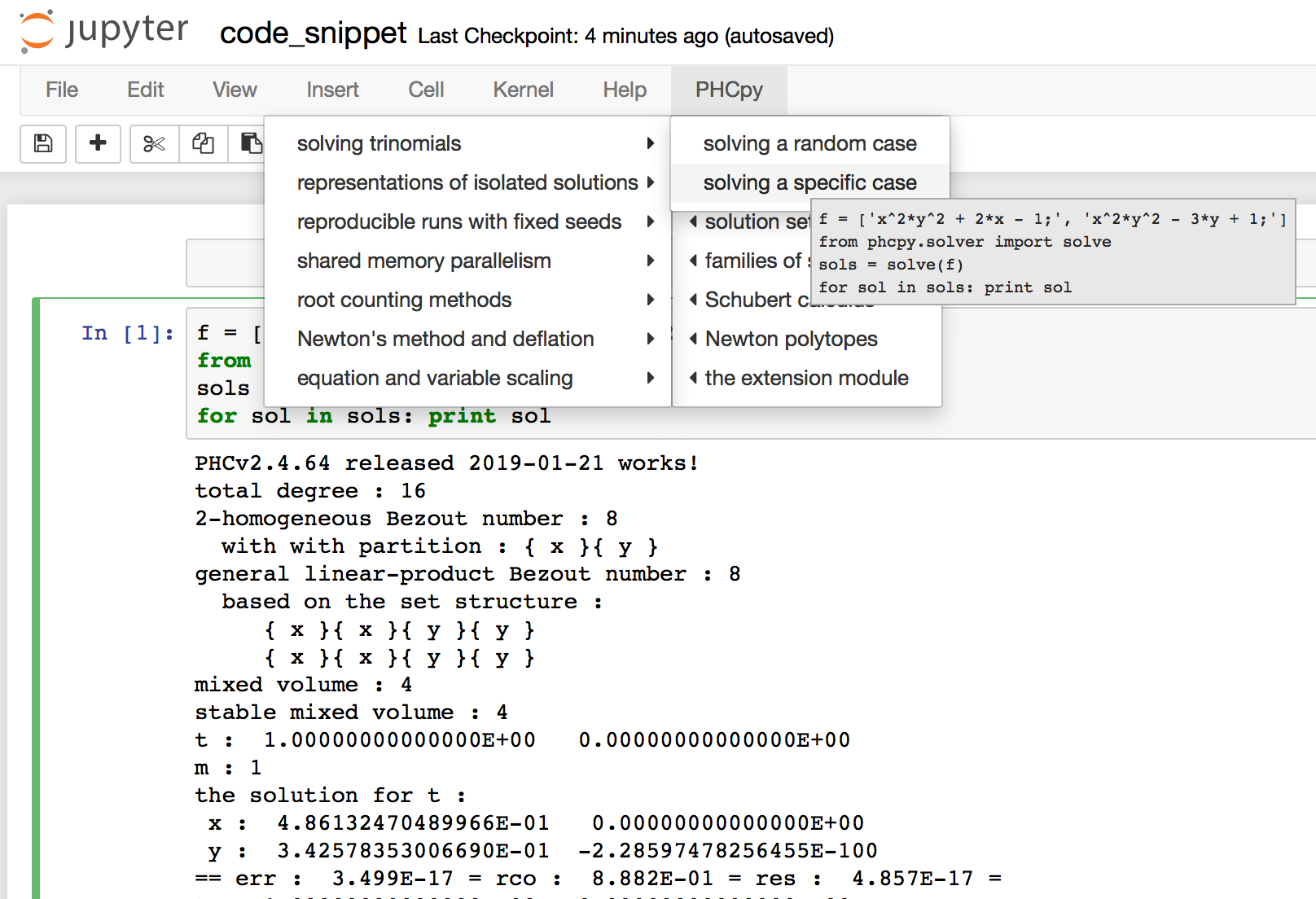}}
\caption{The code snippet for the blackbox solver.  \label{figsnippet}}
\end{figure}

The code snippets from these tutorials are available in our JupyterHub 
deployment, via the snippets menu provided by nbextensions \cite{JUP15}. 
This menu suggests typical applications to guide the novice user. 
The screen shot in Fig.~\ref{figsnippet} 
shows the code snippet reproduced below.

\vspace{1mm}
\begin{verbatim}
   # PHCpy > blackbox solver > solving trinomials
   #       > solving a specific case
   from phcpy.solver import solve

   f = ['x^2*y^2 + 2*x - 1;', 'x^2*y^2 - 3*y + 1;']
   sols = solve(f)
   for sol in sols: print(sol)
\end{verbatim}
\vspace{1mm}
The first solution of the given trinomial can be read 
as (0.48613... + 0.0i, 0.34258... - 0.0i), 
where the imaginary part of x is exactly zero, and that 
of y negligibly small. 
Programmatically, these can be accessed using either 
\texttt{solve(f, dictionary\_output=True)}, 
or equivalently by parsing strings through 
\newline \texttt{{[}phcpy.solutions.strsol2dict(sol) for sol in solve(f){]}}.

\subsection{Direct Manipulation}

One consequence of the Jupyter notebook's rich output is the possibility 
of rich input, as explored through ipywidgets \cite{IPYW} and interactive 
plotting libraries. The combination of rich input with fast numerical 
methods makes surprising interactions possible, such as interactive 
solution of Apollonius' Problem, which is to construct all circles tangent 
to three given circles in a plane.

The tutorial given in the phcpy documentation was adapted for a demo 
accompanying a SciPy poster in 2017, whose code \cite{APP} will run on 
our JupyterHub (by copying \texttt{apollonius\_d3.ipynb} 
and \texttt{apollonius\_d3.js} to one's own user directory).

\begin{figure}[h]
{\includegraphics[width=\columnwidth]{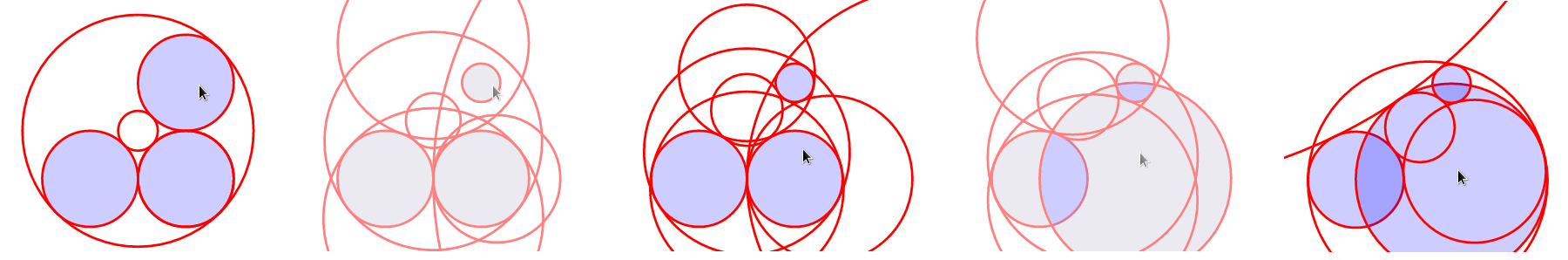}}
\caption{Tangent circles calculated by phcpy in response to user 
reparameterization of the system. \label{apollonius}}
\end{figure}

This system of 3 nonlinear constraints in 5 parameters for each 
of 8 possible tangent circles can be solved interactively by our system 
in real-time (Fig.~\ref{apollonius}).  Although any of the 8 tangent circles
could have nonzero imaginary part in their x/y position or radius, depending
on input coefficients (input circles), such circles are not rendered. 
Thanks to its rich output capabilities, Jupyter is a suitable environment for
mapping algebraic inputs to the planar geometric objects they represent 
(a data binding) through D3.js~\cite{D3}.

This approach makes use of the real-time solution of small polynomial systems,
demonstrating the low latency of phcpy.
It complements static input conditions by investigating their continous
deformation, especially across singular solutions (which PHCpack handles
more robustly than naive homotopy methods).
Singular solutions of polynomial systems are handled by deflation \cite{LVZ06},
which restores quadratic convergence of Newton's method by the addition of
sufficiently many higher order derivatives to the original system.

\section{Solving Polynomial Systems}

Our input is a list of polynomials in several variables.
This input list represents a polynomial system.
By default, the coefficients of the polynomials are considered
as complex floating point numbers.
The system is then solved over the field of complex numbers.

For general polynomial systems,
the complexity of the solution set can be expected to grow
exponentially in the dimensions
(number of polynomials and variables) of the system.
The complexity of computing all solutions of a polynomial system is \#P-hard.
The complexity class \#P is the class of counting problems.
Formulating instances of polynomial systems that will occupy
fast computers for a long time is not hard.

\subsection{Polynomial Homotopy Continuation}

By computing over the field of complex numbers, we exploit the continuity
of the solution set as a function of the coefficients of the polynomials 
in the system.
These numerical algorithms, called continuation methods, 
track solution paths defined by
a one parameter family of polynomial systems (the homotopy). 
Homotopy methods take a polynomial system as input, and construct 
a suitable embedding of the input system
into a family which contains a start system with known solutions.

We say that a homotopy is \emph{optimal} if for generic instances of
the coefficients of the input system no solution paths diverge.
Even as the complexity of the solution set is very hard,
the problem of computing the next solution, or just one random solution,
has a much lower complexity.  phcpy offers optimal homotopies for
three classes of polynomial systems:

\begin{enumerate}

\item \emph{dense polynomial systems}

A polynomial of degree \emph{d} can be deformed into a product of \emph{d}
linear polynomials.
If we do this for all polynomials in the system (as in \cite{VC93}),
then the solutions of the deformed system are solutions of linear systems.
Continuation methods track the paths originating at the solutions of
the deformed system to the given problem.

\item \emph{sparse polynomial systems}

A system is sparse if relatively few monomials appear with nonzero
coefficient.  The convex hulls of the exponent vectors of the monomials
that appear are called Newton polytopes.  The mixed volume of the
tuple of Newton polytopes associated with the system is a sharp upper
bound for the number of isolated solutions.
Polyhedral homotopies (\cite{HS95}, \cite{VVC94})
start at solutions of systems that are sparser than the given system
and extend those solutions to the solutions of the given problem.

\item \emph{Schubert problems in enumerative geometry}

The classical example is to compute all lines in 3-space that
meet four given lines nontrivially.
Homotopies to solve geometric problems move the input data
to special position, solve the special configuration, and then
deform the solutions of the special problem into those of the
original problem.  Such homotopies were introduced in \cite{HSS98}.

\end{enumerate}

All classes of homotopies share the introduction of random constants.

For its fast mixed volume computation,
the software incorporates MixedVol \cite{GLW05} and DEMiCs \cite{MT08}.
High-precision double double and quad double arithmetic is performed
by the algorithms in QDlib \cite{HLB01}.

\subsection{Speedup and Quality Up}

The solution paths defined by polynomial homotopies can be tracked
independently, providing obvious opportunities for parallel execution.
This section reports on computations on our server, a 44-core computer.

An obvious benefit of running on many cores is the speedup.
The \emph{quality up} question asks the following: if we can afford to spend
the same time, by how much can we improve the solution 
using \emph{p} processors?

We illustrate the quality up question on the cyclic 7-roots
benchmark problem \cite{BF91}.
The online SymPy documentation \cite{SymPyDocs} 
uses the cyclic 4-roots problem
to illustrate its \texttt{nonlinsolve} method.

The function defined below returns the elapsed performance
of the blackbox solver on the cyclic 7-roots benchmark problem,
for a number of tasks and a precision equal to double, double double,
or quad double arithmetic.

\begin{verbatim}
    def qualityup(nbtasks=0, precflag='d'):
        """
        Runs the blackbox solver on a system.
        The default uses no tasks and no multiprecision.
        The elapsed performance is returned.
        """
        from phcpy.families import cyclic
        from phcpy.solver import solve
        from time import perf_counter
        c7 = cyclic(7)
        tstart = perf_counter()
        s = solve(c7, verbose=False, tasks=nbtasks, \
                  precision=precflag, checkin=False)
        return perf_counter() - tstart
\end{verbatim}

\vspace{1mm}

The function above is applied in an interactive Python script,
prompting the user for the number of tasks and precision,
This scripts runs in a Terminal window and prints the elapsed performance
returned by the function.
If the quality of the solutions is defined as the working precision,
then to answer the quality up question,
one considers how many processors are needed
to compensate for the overhead of the multiprecision arithmetic.

Although cyclic 7-roots is a small system for modern computers,
the cost of tracking all solution paths in double double and
quad double arithmetic causes significant overhead.
The script above was executed on a 2.2 GHz Intel Xeon E5-2699 processor
in a CentOS Linux workstation with 256 GB RAM
and the elapsed performance is in Table~\ref{perfcyc7overhead}.

\begin{table}[hbt]
\begin{center}
\begin{tabular}{l|r|r|r}
precision & d~~ & dd~~ & qd~~ \\ \hline
elapsed performance & 5.45 & 42.41 & 604.91 \\
overhead factor & 1.00 & 7.41 & 110.99
\end{tabular}
\caption{Elapsed performance of the blackbox solver in double,
        double double, and quad double precision. \label{perfcyc7overhead}}
\end{center}
\end{table}

Table~\ref{perfcyc7parallel} demonstrates the reduction of the
overhead caused by the multiprecision arithmetic by multitasking.

\begin{table}[hbt]
\begin{center}
\begin{tabular}{c|r|r|r}
tasks & 8~~ & 16~~ & 32~~ \\ \hline
dd & 7.56 & 5.07 & 3.88 \\
qd & 96.08 & 65.82 & 44.35 
\end{tabular}
\caption{Elapsed performance of the blackbox solver
        with 8, 16, and 32 path tracking tasks, in double double
        and quad double precision. \label{perfcyc7parallel}}
\end{center}
\end{table}

Notice that the 5.07 in Table~\ref{perfcyc7parallel}
is less than the 5.45 of Table~\ref{perfcyc7overhead}:
with 16 tasks we doubled the precision and finished the computations
in about the same time.
The 42.41 and 44.35 in Table~\ref{perfcyc7parallel} are similar enough
to state that with 32 instead of 1 task we doubled the precision from
double double to quad double precision in about the same time.

\begin{figure}[h]
{\includegraphics[width=\columnwidth]{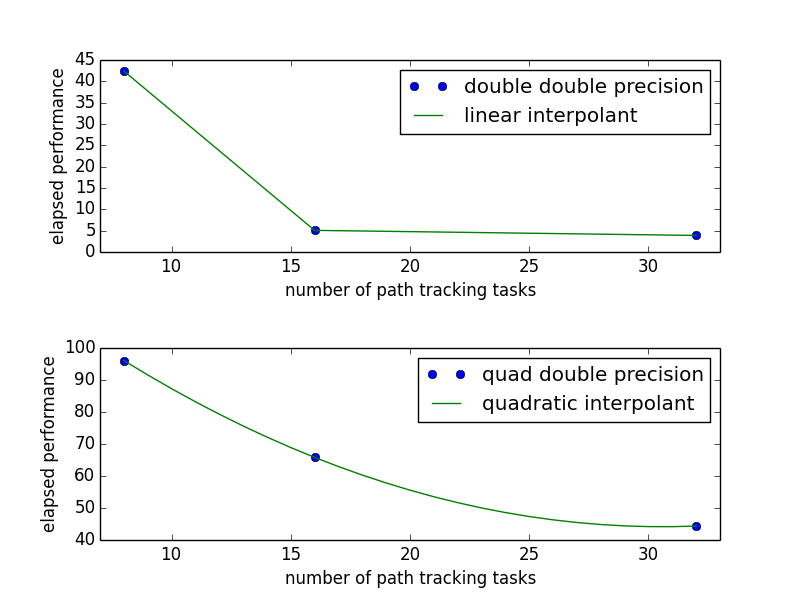}}
\caption{Interpolated elapsed performances.  \label{figqualityup}}
\end{figure}

The data in Table~\ref{perfcyc7parallel} is
visualized in Fig.~\ref{figqualityup}.
The interpolation allows us to estimate running times for a number
of tasks different from the measured run times.
To answer the original quality up question,
one could interpolate between the sizes of working precision
to answer the following quality up question.
If we can afford to spend the same time as on one path tracking task,
then how many extra decimal places can we gain with \emph{p} 
path tracking tasks?

Precision is a crude measure of quality.
Another motivation for quality up by parallelism is to compensate
for the cost overhead caused by arithmetic with power series.
Power series are hybrid symbolic-numeric representations
for algebraic curves.

\subsection{Positive Dimensional Solution Sets}

\emph{Solving} a system has evolved in meaning, 
from computing approximations of all its isolated solutions, 
to finding the numerical irreducible decomposition
of the solution set.  The numerical irreducible decomposition includes
not only the isolated solutions, but also the representations for all
positive dimensional solution sets. Such representations consist
of sets of \emph{generic points}, partitioned along the irreducible factors.

To illustrate this expanded sense of a 'solution', 
we consider the twisted cubic,
known in algebraic geometry as the first nontrivial space curve.
We use this example to illustrate two different representations
of this space curve:

\begin{enumerate}

\item 
In a \emph{witness set} construction, the given polynomial equations are
augmented with as many generic hyperplanes as the dimension of the
solution set.  The solutions which satisfy the system and the augmented
equations are generic points.  As the degree of the twisted cubic is
three, we find three points on a random plane intersecting the cubic.

\vspace{1mm}

\begin{verbatim}
      pols = ['x*y - z;', 'x^2 - y;']
      from phcpy.sets import embed
      from phcpy.solver import solve
      embp = embed(3, 1, pols)
      sols = solve(embp, verbose=False)
      print('#generic points :', len(sols))
\end{verbatim}

\vspace{1mm}

The above snippet constructs the embedding for the equations that
define the twisted cubic.
The solutions of this embedding represent the curve.
Moving the added plane and tracking the solution paths starting at
the three generic points will provide many more samples of the curve.

\item 
A \emph{series expansion} for the solution starts its development at
some point(s) in a coordinate hyperplane.
In this hyperplane, the curve intersects the solution set at some point(s).
For a simple example as the twisted cubic, the series development
defines an exact solution after the initial term.
Consider the snippet:

\vspace{1mm}

\begin{verbatim}
      pols = ['x*y - z;', 'x^2 - y;']
      from phcpy.maps import solve_binomials
      maps = solve_binomials(3, pols, \
                 puretopdim=True)
      for sol in maps:
          print(sol)
\end{verbatim}

\vspace{1mm}

\noindent The output of the above snippet is

\vspace{1mm}

\begin{verbatim}
      ['x - (1+0j)*t1**1', 'y - (1+0j)*t1**2', \
       'z - (1+0j)*t1**3', 'dimension = 1', \
       'degree = 3']
\end{verbatim}

\vspace{1mm}

which corresponds to the parametric respresentation
$(t, t^2, t^3)$ of the twisted cubic.
\end{enumerate}

Many interesting polynomial systems have isolated solutions
and positive dimensional solution sets.
We consider again the family of cyclic \emph{n}-roots problems,
now for $n = 8$, \cite{BF94}.
While for $n = 7$ all roots are isolated points,
there is a one dimensional solution curve of cyclic 8-roots of degree 144.
This curve decomposes in 16 irreducible factors,
eight factors of degree 16 and eight quadratic factors,
adding up to $8 \times 16 + 8 \times 2 = 144$.

Consider the following code snippet.

\vspace{1mm}

\begin{verbatim}
    from phcpy.phcpy2c3 import py2c_set_seed
    from phcpy.factor import solve
    from phcpy.families import cyclic
    py2c_set_seed(201905091)  # for a reproducible run
    c8 = cyclic(8)
    sols = solve(8, 1, c8, verbose=False)
    witpols, witsols, factors = sols[1]
    deg = len(witsols)
    print('degree of solution set at dimension 1 :', deg)
    print('number of factors : ', len(factors))
    _, isosols = sols[0]
    print('number of isolated solutions :', len(isosols))
\end{verbatim}

\vspace{1mm}

\noindent The output of the script is

\vspace{1mm}

\begin{verbatim}
    degree of solution set at dimension 1 : 144
    number of factors :  16
    number of isolated solutions : 1152
\end{verbatim}
This numerical output is the essence of the blackbox solver
for positive dimensional solution sets \cite{Ver18}.

\section{Survey of Applications}

We consider some examples from various literatures 
which apply polynomial constraint solving. The first two examples 
use phcpy in particular as a research tool. 
The remaining three are broader examples representing current 
uses of numerical algebraic geometry in other STEM fields.

\subsection{Rigid Graph Theory}

The conformations of proteins \cite{LML14}, molecules \cite{EM99}, 
and robotic mechanisms (discussed further below) can be studied by 
counting and classifying unique mechanisms, i.e. real embeddings 
of graphs with fixed edge lengths, modulo rigid motions, 
per Bartzos et al. \cite{BELT18}.

Consider a graph $G$ whose edges $e \in E_G$ each have a given 
length $d_{e}$. A graph embedding is a function that maps the vertices 
of $G$ into $D$-dimensional Euclidean space (especially $D$ = 2 or 3). 
Embeddings which are 'compatible' are those which preserve $G$'s edge lengths.
The number of unique mechanisms is thus a function of $G$ and $\mathbf{d}$.
An upper bound over all $d$ and $G$ with k vertices (yielding lower bounds 
for graphs with $n \geq k$ vertices, unless the upper bound is infinite) 
can be computed. In particular, the Cayley-Menger matrix of $\mathbf{d}$ 
\cite{LLMM14} (i.e., the squared distance matrix with a row and column 
of 1s prepended, except that its main diagonal is 0s) is an algebraic system,
proportional to the mixed volume.   Certain of its square subsystems 
characterize the mechanism in terms of these bounds on unique mechanisms.

Bartzos et al. implemented, using \texttt{phcpy}, a constructive method 
yielding all 7-vertex minimally rigid graphs in 3D space (the smallest open
case) and certain 8-vertex cases previously uncounted.
A graph $G$ is generically rigid if, for any given edge lengths $d$,
none of its compatible embeddings (into a generic configuration such
that vertices are algebraically independent) are continuously deformable.
$G$ is minimally rigid if removing any one of its edges 
yields a non-rigid mechanism.

\texttt{phcpy} was used to find edge lengths with maximally many real 
embeddings, exploiting the flexibility of being able to specify their 
starting system.  This sped up their algorithm by perturbing the solutions
of previous systems to find new ones.

Many iterations of sampling have to be performed if the wrong number of 
real embeddings is found; in each case, a different subgraph is selected
based on a heuristic implemented by the \texttt{DBSCAN} class 
of \texttt{scikit-learn} (illustrating the value of a scientific Python 
ecosystem).  The actual number of real embeddings is known from an 
enumeration of unique graphs constructed by Henneberg steps in, 
for instance, SageMath.

\subsection{Model Selection \& Parameter Inference}

It is often useful to know all the steady states of a biological network, 
as represented by a nonlinear system of ordinary differential equations, 
with some conserved quantities. These two lists 
of polynomials (from rates of change of form $\dot{x} = p(x)$, 
by letting $\dot{x}=0$; and from conservation laws of form $c = \sum{x_i}$
by subtracting $c$ from both sides) have a zero set which is a steady-state 
variety, that can be explored numerically via polynomial homotopy continuation.

Parameter homotopies were used by Gross et al. \cite{GHR16} to perform model 
selection on a mammalian phosphorylation pathway, determining whether the 
kinase acts processively (i.e. adding more than one phosphate at once,
which it does not in vitro). Their analysis validated experimental work
showing processivity in vivo. In doing so, they obtained >50x speedup over
non-parameter homotopies (for running times in minutes, not hours) on systems
tracking 20 paths.

\subsection{Critical Point Computation}

Polynomial homotopy continuation has also been adapted to the field of 
chemical engineering to locate critical points of multicomponent mixtures
\cite{SWM16}, i.e., temperature and pressure satisfying a multi-phase
equilibrium.

A remarkable variety of systems of constraint also take on polynomial form,
or can be approximated thereby, in various sciences. 
Diverse problems in the analysis of belief propagation (in graphical models)
\cite{KMC18}, hyperbolic conservation laws (in PDEs) \cite{HHS13}, 
and vacuum moduli spaces (in supersymmetric field theory) \cite{HHM13} 
have been addressed using polynomial homotopy continuation.

\subsection{Algebraic Kinematics}

We have discussed an application of numerical methods to count
ing unique instances of rigid-body mechanisms. In fact, kinematics 
and numerical algebraic geometry have a close historical relationship.
Following Wampler and Sommese \cite{WS11}, other geometric problems arising 
from robotics include \textbf{analysis} of specific mechanisms e.g.,:%

\begin{itemize}

\item Motion decomposition - into assembly modes (of individual mechanisms)
or subfamilies of mechanisms (with varying mobility);

\item Mobility analysis - degrees of freedom of a mechanism
(sometimes exceptional),
sometimes specific to certain configurations (e.g., gimbal lock);

\item Kinematics - effector position given parameters (forward kinematics),
and vice versa (inverse kinematics, e.g. used in computer animation);

\item Singularity analysis - detection of situations where the mechanism
can move without change to its parameters (input singularity),
or the parameters can change without movement of the mechanism
(output singularity);

\item Workspace analysis - determining all possible outputs of the mechanism,
i.e.: reachable poses;
\end{itemize}
...as well as the \textbf{synthesis} of mechanisms that can reach certain sets
of outputs, or that can be controlled by a certain input/output relationship.

\begin{figure}[h]
{\includegraphics[width=\columnwidth]{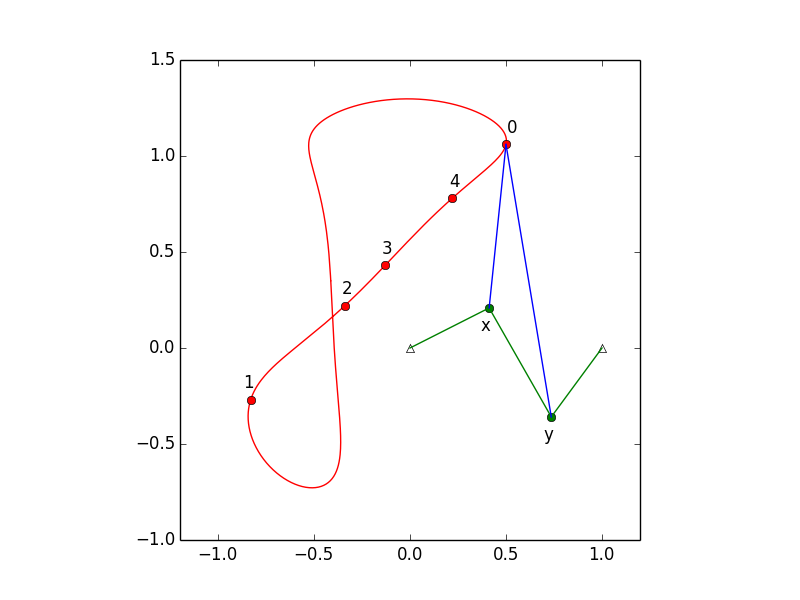}}
\caption{The design of a 4-bar mechanism.  \label{fig4barcoupler}}
\end{figure}

Fig.~\ref{fig4barcoupler} illustrates a reproduction
of one synthesis result in the mechanism design literature \cite{MW90}.
Given five points, the problem is to determine the length of two bars
so their coupler curve passes through the five given points.

This example is part of the tutorial of phcpy and the scripts
to reproduce the results are in its source code distribution.
The equations are generated with sympy~\cite{SymPy}
and the plots are made with matplotlib~\cite{Hun07}.

Continuation homotopies were developed as a substitute for algebraic 
elimination that was more robust 
to special cases, yet still tractable to numerical techniques. 
Research in kinematics increasingly relies on such algorithms~\cite{WS11}.

\subsection{Systems Biology}

Whether a model biological system is multistationary or oscillatory, 
and whether this depends on its rate constants, are all properties 
of its steady-state locus. Following the survey of 
Gross et al. \cite{GBH16} regarding uses of numerical algebraic geometry 
in this domain, one might seek to:

\begin{itemize}
\item determine which values of the rate and conserved-quantity parameters 
      allow the model to have multiple steady states;

\item evaluate models with partial data (subsets of the $x_i$) and reject
      those which don't agree with the data at steady state;

\item describe all the states accessible from a given state of the model,
      i.e. that state's stoichiometric compatibility class (or basin 
      of attraction);

\item determine whether rate parameters of the given model are identifiable
      from concentration measurements, or at least constrained.
\end{itemize}

For large real-world models in systems biology, these questions of algebraic 
geometry are only tractable to numerical methods scaling to many dozens 
of simultaneous equations.

\section{Conclusion}

From these examples, we see that polynomial homotopy continuation 
has wide applicability to STEM fields. 
Moreover, phcpy is an accessible interface to the technique, 
capable of high performance whilst producing certifiable and
reproducible results.

\subsection{Acknowledgments}

This material is based upon work supported
by the National Science Foundation under Grant No.~1440534.

\end{document}